\documentclass[12pt]{article}

\usepackage{scicite}

\usepackage{times}
\usepackage{xcolor}
\usepackage{graphicx}
\usepackage{ulem}

\newenvironment{sciabstract}{%
\begin{quote} \bf}
{\end{quote}}

\title{Quantum sensing of photonic spin density} 

\author
{Farid Kalhor$^{1,*}$, Li-Ping Yang$^{1,2,*}$, Leif Bauer$^{1}$, Zubin Jacob$^{1 \dag}$\\
\\
\normalsize{$^{1}$School of Electrical and Computer Engineering, Purdue University, West
Lafayette, IN 47906, USA,}\\
\normalsize{$^{2}$Center for Quantum Sciences and School of Physics, Northeast Normal University,} \\ \normalsize{Changchun 130024, China}\\

\\
\normalsize{$^\dag$To whom correspondence should be addressed; E-mail:  zjacob@purdue.edu.}\\
\normalsize{$^*$Authors with equal contribution.}
}

\date{}

\begin{document} 

\baselineskip14pt

\maketitle 

\begin{sciabstract}
  Photonic spin density (PSD) in the near-field gives rise to exotic phenomena such as photonic skyrmions, optical spin-momentum locking and unidirectional topological edge waves. Experimental investigation of these phenomena requires a nanoscale probe that directly interacts with PSD. Here, we propose and demonstrate that the nitrogen-vacancy (NV) center in diamond can be used as a quantum sensor for detecting the spinning nature of photons. This room temperature magnetometer can measure the local polarization of light in ultra-subwavelength volumes through photon-spin-induced virtual transitions. The direct detection of light's spin density at the nanoscale using NV centers in diamond opens a new frontier for studying exotic phases of photons as well as future on-chip applications in spin quantum electrodynamics (sQED). 

\end{sciabstract}

% \section*{Introduction}
The spinning field of light has long been associated with the concept of global polarization \cite{beth_mechanical_1936, raman_experimental_1932}. Here, the spin angular momentum (SAM) of light is a vector with its direction pinned parallel to the momentum of a far field propagating wave. In stark contrast, the photonic spin density (PSD) has only recently emerged to the forefront of nanophotonics \cite{mechelen_universal_2016,bliokh_extraordinary_2014,bliokh_spinorbit_2015,shomroni_all-optical_2014,yin_spin-resolved_2020}. PSD in confined or structured light beyond the traditional paraxial regime can exhibit exotic spatial variation of local polarization known as spin texture. In recent years, exploring the near-field properties of this spin texture has led to the discovery of exotic phenomena such as photonic skyrmions and topological electromagnetic phases of matter \cite{mechelen_photonic_2019,du_deep-subwavelength_2019,tsesses_optical_2018}. 

Striking phenomena originating from near-field photon spin density include directional spontaneous emission, one-way scattering of surface plasmon polaritons, transverse spin in free space light beams, and anomalous optical forces \cite{kalhor_universal_2016,zhang_all-optical_2017,rodriguez-fortuno_lateral_2015,neugebauer_measuring_2015,javadi_spinphoton_2018}. Here, the nature of PSD is inferred indirectly through directional phenomena i.e.  spin-momentum locking or spin to orbital angular momentum conversion \cite{sayrin_nanophotonic_2015,arzola_spin_2019,gong_nanoscale_2018}. We note that the orbital angular momentum (OAM) of light is an extrinsic degree of freedom that can be directly detected due to its wavelength scale phase signatures \cite{ji_photocurrent_2020}.
However, direct measurement of PSD remains a challenge since the fundamental property of photon spin density exists in ultra-subwavelength volumes of the light field. Therefore, there is an urgent need to develop a nanoscale photon spin probe similar to near-field scanning optical microscopy \cite{neugebauer_measuring_2015, yin_spin-resolved_2020} routinely used to detect dipolar electric fields or magnetic force microscopy which can map nanoscale texture of electron spins.

Here, we propose and demonstrate that nitrogen vacancy centers in diamond can sense the local polarization of spinning light fields within ultra small mode volumes. Our work offers a paradigm shift in the quantum sensing of photon spin since NV centers in diamond have only been used to probe electron spin excitations in matter such as magnons \cite{van_der_sar_nanometre-scale_2015}, magnetic thin films \cite{thiel_probing_2019}, and magnetic skyrmions \cite{dovzhenko_magnetostatic_2018}. Our sensor  functions through the effective static magnetic field generated by PSD on interaction with a single spin qubit. We measure the induced phase of the spin qubit through optical read out to directly detect the PSD of a detuned laser beam that only causes virtual transitions. We demonstrate coherent interaction at room temperature paving the way to probe exotic spin states of photons. Finally, we shed light on how on-chip nanophotonic structures possess effective magnetic fields arising from the intrinsic spin of evanescent waves. Our discovered nanophotonic phenomenon can be exploited in future spin quantum electrodynamics (sQED) devices for on-chip and targeted addressing of spin qubits.

\begin{figure}
    \centering
    \includegraphics[width=120mm]{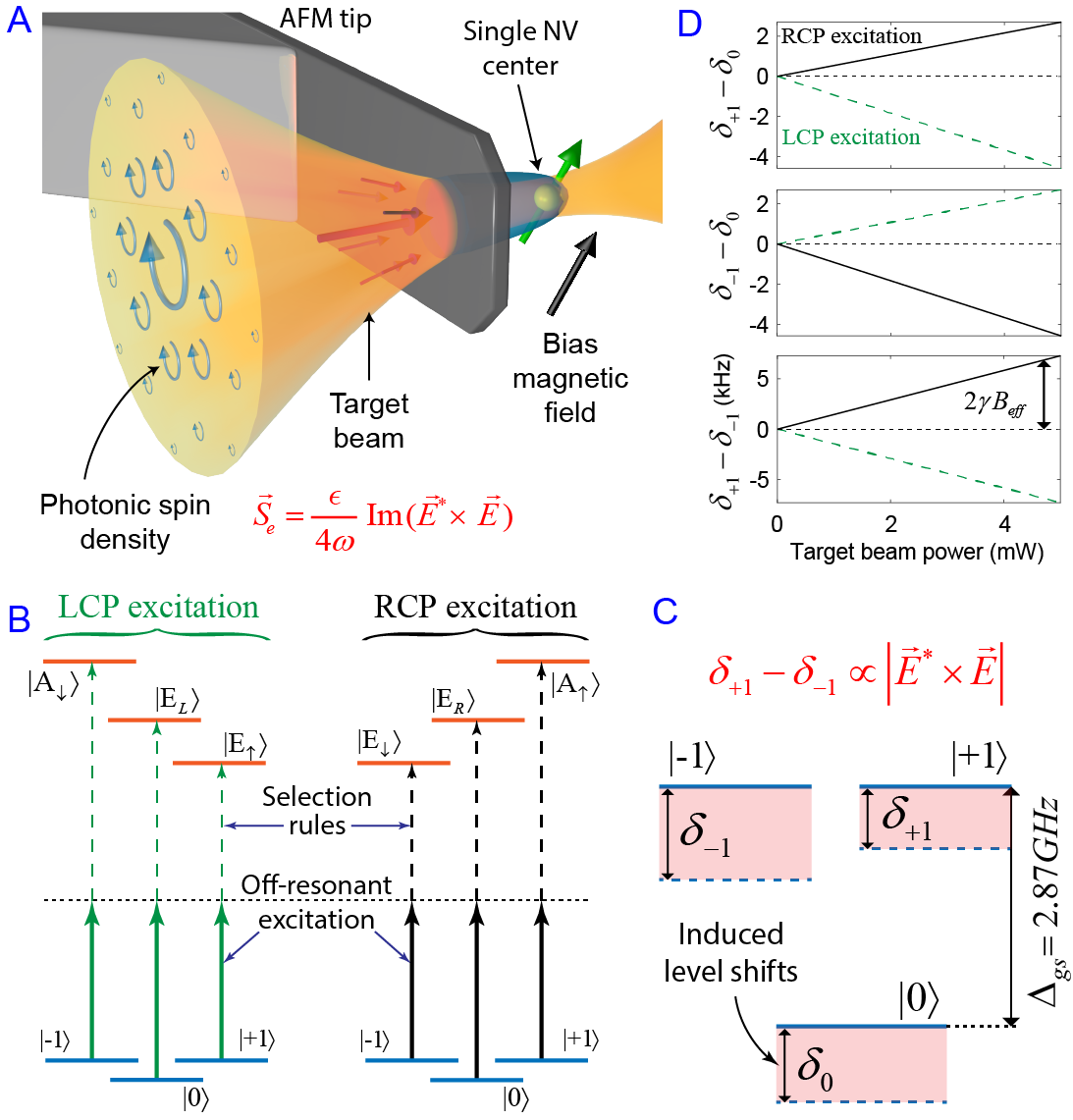}
    \caption{Probing photonic spin density (PSD) with a single NV center. (A)  A target beam, red-detuned to the NV center transition, is incident on a single NV center on an AFM tip. We measure the relative energy level shifts in the ground state and show its proportionality to the PSD. The single NV center serves as a room temperature nanoscale probe for PSD. (B) Ground and excited state energy levels of an NV center showing the selection rules for RCP and LCP excitation. (C) Level shifts induced in the ground state due to the virtual transitions under the off-resonant target beam. (D) Power and polarization (spin) dependence of the relative energy shifts in the ground state, resembling Zeeman splitting. An effective magnetic field is defined as $B_{eff}=(\delta_{+1}-\delta_{-1})/2\gamma$.}
    \label{fig1}
\end{figure}

In order to demonstrate this ultra-subwavelength probing of PSD, we study the interaction between the spin of a monochromatic optical beam and an NV center that is placed on an atomic force microscopy (AFM) tip. The directly observable part of the PSD of a laser beam is given by $\vec{S}^{\rm obs}_E=\epsilon \vec{E}_{\perp}(\vec{r},t)\times \vec{A}_{\perp}(\vec{r},t)$~\cite{yang2020quantum}, where $\vec{E}$ is the electric field, $\vec{A}$ is the vector potential, the subindex $\ _{\perp}$ denotes the transverse part of the vector field, and $\epsilon$ is the permittivity. In the monochromatic single-frequency limit, the PSD of a beam with frequency $\omega$ can be rewritten as $\vec{S}^{\rm obs}_E=-(i\epsilon/4\omega)\vec{E}^*\times\vec{E}$  where $\vec{E}^{*}$ denotes the complex conjugate of the complex electric field \cite{berry_optical_2009,picardi_angular_2018}. The PSD is time-independent and is related to the handedness of the polarization of the beam. According to the selection rules of electric-dipole transitions, circularly polarized light will change the electronic orbital angular momentum by $\pm\hbar$ while keeping the electron-spin state unchanged. However, due to  spin-orbit coupling, the transition frequencies in the NV center become dependent on the electron spin states as shown in Fig.~\ref{fig1}B. Under a detuned incident light beam, virtual electric-dipole transitions will induce AC Stark shifts in the ground electronic state~\cite{Deutsch1998quantum,Buckley2010spinlight,albrecht_fictitious_2016,wilkinson_spin-selective_2019}.
As illustrated in Fig. 1C\&D, the amplitude of these shifts ($\delta_0$ and $\delta_{\pm 1}$) depend on both the electronic spin state of the NV center and the photonic spin density of the excitation. We show that this effect manifests itself as a PSD dependent effective magnetic field. 
We exploit the single NV center as a nanoscale quantum magnetometer \cite{Degen2017Quantum} to measure this effective static magnetic field created by the target spinning light.

We consider the detailed energy level structure of NV centers in the basis of RCP and LCP transitions to express the net energy shift as a function of PSD. Using second order perturbation theory for the basis transformed NV center Hamiltonian (see supplementary information), we arrive at the result of the effective magnetic field generated by spinning light: 
\begin{equation}
B_{eff}\equiv\frac{\delta_{+1}-\delta_{-1}}{2\gamma}\approx \frac{C}{\lambda_0\Delta^2}(\vec{S}_E^{obs}\cdot\hat{n}),\label{eq:B_eff}
\end{equation}
where $\gamma\approx 2.8$~MHz/G is the gyro-magnetic ratio of the NV spin, $C$ is a constant determined by the intrinsic properties of the NV, $\lambda_0$ is the wavelength of the off-resonant excitation, $\Delta$ is the detuning between the frequency of the target light and the optical transition of the NV center, and $\hat{n}$ is the direction of the NV center (see supplementary information A). From Eq. 1, we see that the strength of the effective magnetic field is proportional to the projection of the PSD ($\vec{S}_E^{obs}$) on the NV center axis ($\hat{n}$). 
In the expression above, we have used the spin states $|\pm1\rangle$ to form the probe qubit as it leads to effective static magnetic fields directly proportional to the PSD. One can also use $|0\rangle\leftrightarrow|\pm1\rangle$ transitions as the probe qubit. We show the detailed comparison between the choice of probe qubits in Fig. 2C.

\begin{figure}
    \centering
    \includegraphics[width=120mm]{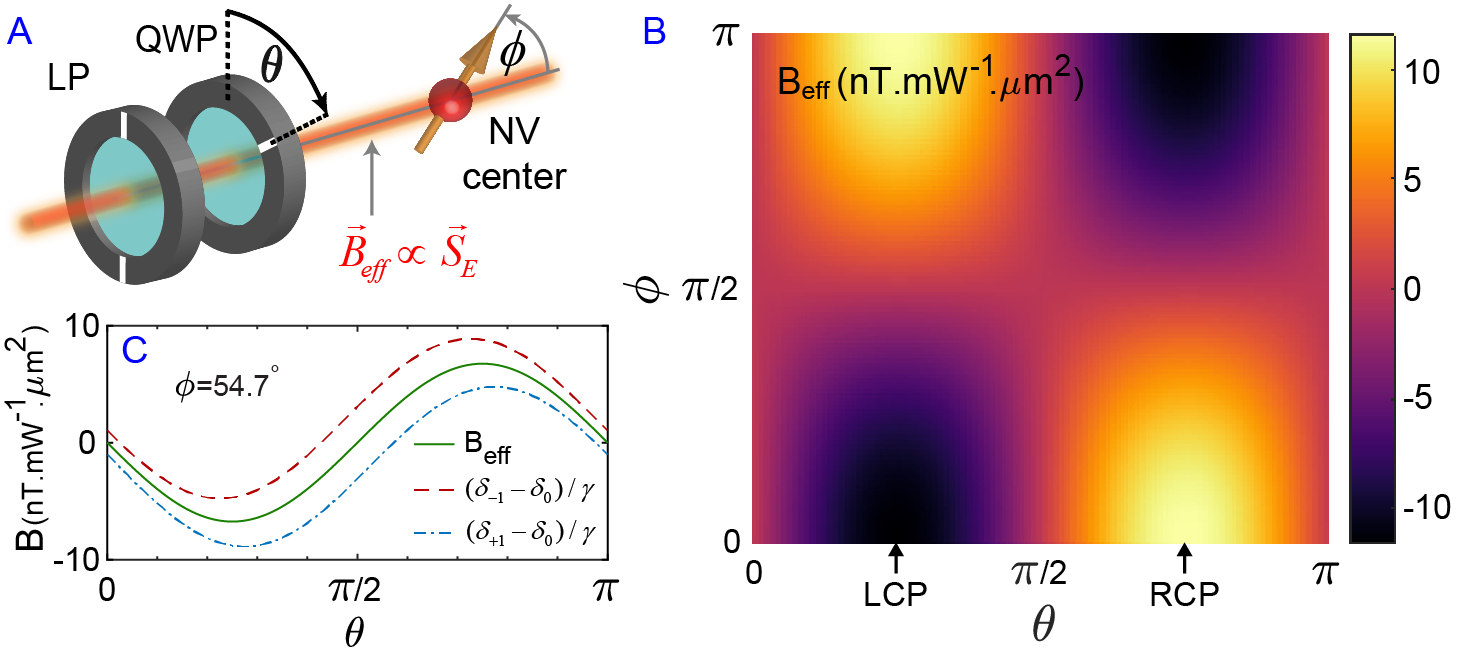}
    \caption{Effective magnetic field induced by PSD. (A) A linear polarizer and a quarter wave plate control the PSD. (B) The effective field calculated for different QWP angles ($\theta$) and alignment angles ($\phi$). (C) The effective field calculated for an NV center in a (100) cut diamond, $\phi=54.7^\circ$.}
    \label{fig2}
\end{figure}

The PSD of the target beam is determined by its power and the average spin of each photon. In our experiment, the wavelength of the target beam is $\lambda_0=800nm$, which is far off-resonant with the optical transition of the NV center at $\lambda=637nm$ (Fig.~\ref{fig1}B).  We control the PSD by tuning the angle $\theta$ between the linear polarizer and the quarter-wave plate (QWP) (see Fig. 2A). The degree of circular polarization (i.e., the photonic spin density) is $\vec{S}_E^{obs}\propto sin(2\theta)\hat{z}$, where $\hat{z}$ is the direction of propagation of the beam. An NV center can only sense the projection of a magnetic field on its defect axis. Therefore, the measured effective field also depends on the alignment angle $\phi$ between the NV center axis and the PSD vector. We show the theoretical simulation of the effective field $B_{eff}$ sensed by an NV center, as a function of $\theta$ and $\phi$ in Fig.~\ref{fig2}B. In our experiment, the alignment angle is fixed at $\phi=54.7^\circ$. For this specific angle, we show the effective magnetic field experienced by the probe qubit in Fig.~\ref{fig2}C. This variation of the effective magnetic field with degree of circular polarization is the unique signature of PSD. We note that the target laser is red-detuned to the optical transition of the NV center and is not absorbed by the NV center. Therefore, the resulting effective field is not due to the absorption or emission related spectral features of the NV center. It is related to the induced phase in the spin qubit measured by optical read out.

\begin{figure}
    \centering
    \includegraphics[width=160mm]{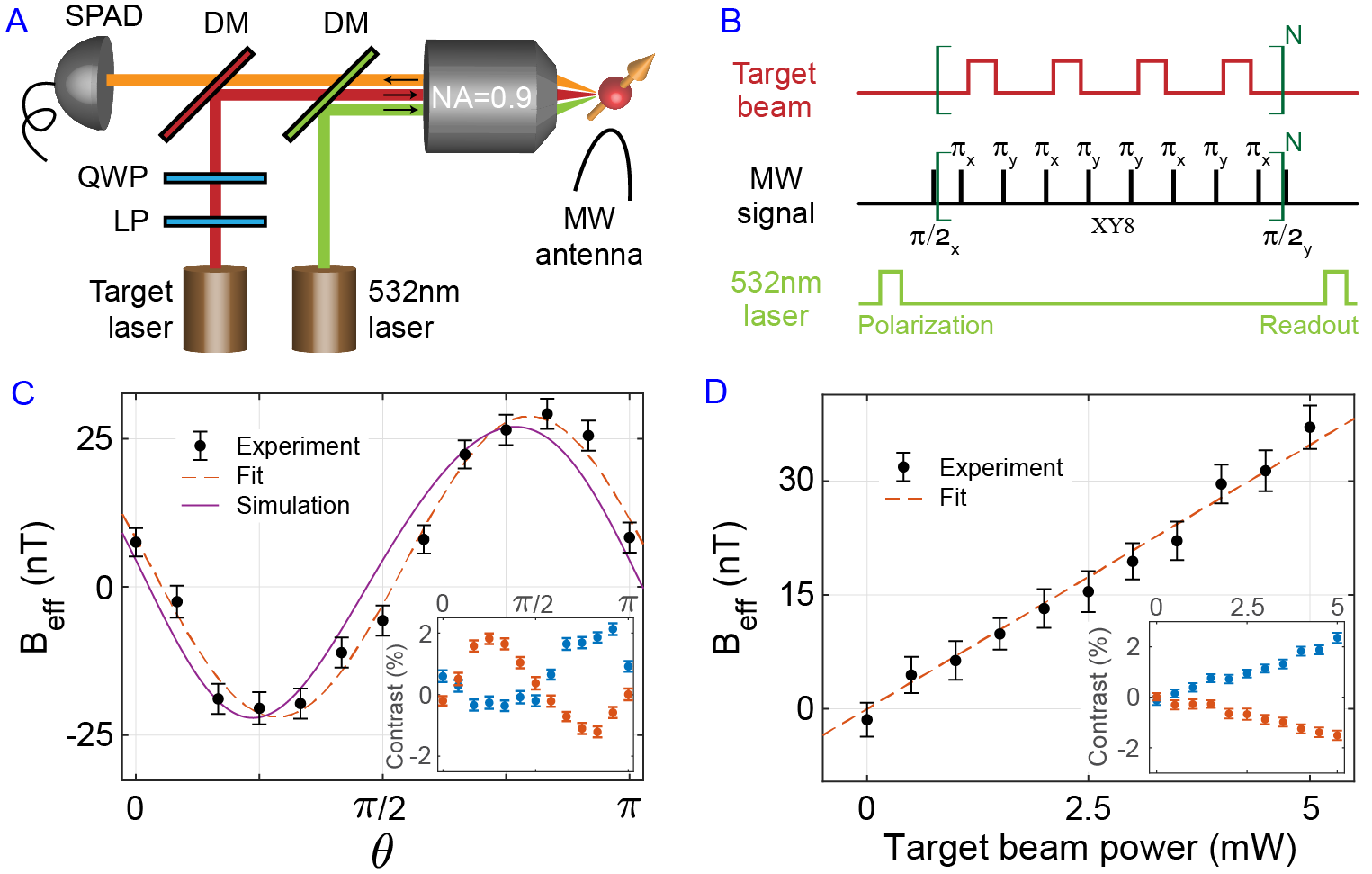}
    \caption{Demonstration of nanoscale PSD probe with a single NV center. (A) Simplified schematic of the experimental setup. (B) Pulse sequence showing dynamical decoupling for AC magnetometry and amplitude modulation of the target beam to generate an AC effective magnetic field suitable for high sensitivity measurement. (C) Measured PSD for different QWP angles with target beam power of 4mW. The dashed red curve is a sine fit to the data, the solid purple curve is the numerical simulation results, matching closely to the measurements. (D) Measured PSD as a function of incident power for $\theta=3\pi/4$ showing a linear dependence. The dashed red curve is a linear fit. Insets show the raw measurement data for $|0\rangle \rightarrow |+1\rangle$ (red) and $|0\rangle \rightarrow |-1\rangle$ (blue) transitions. C\&D show that the effective field is directly proportional to PSD.}
    \label{fig3}
\end{figure}

We overcome the challenge of room temperature observation of PSD to pave the way for future on-chip applications. We exploit a large detuning of the target PSD beam to the optical transition of the NV center  to avoid absorption of photons by the NV center and ensure optimal collection of the emitted PL.
This detuning limits the amplitude of the generated effective static magnetic field to a few tens of nanotesla. In order to probe this effective magnetic field at the location of the single NV center, we leverage AC magnetometry techniques at room temperature. In AC magnetometry, high sensitivity is achieved due to a long coherence time from spectral filtering of magnetic fluctuations (e.g., nuclear noise) coupled to the NV center \cite{doherty_nitrogen-vacancy_2013,taylor_high-sensitivity_2008}. Fig. 3A\&B show simplified schematics of the experiment and dynamics of the measurements (see supplementary information B \& C for more details). A 532nm laser is used to initialize and readout the state of the qubit. After initialization, a  series of microwave pulses are sent in XY8 configuration to achieve dynamical decoupling of the qubit from background noise \cite{ali_ahmed_robustness_2013}. The intensity of the target beam is modulated to match the frequency of the XY8 pulse for AC magnetometry. Furthermore, we modulate the phase of the intensity modulation to decouple the signal from systematic noise in our measurement (see supplementary information B \& C for more details). 

The first striking evidence of photonic spin density measurement is shown in Fig 3C. We observe that the effective magnetic field  generated by PSD directly follows the ellipticity of the polarization of the target beam.  For a Gaussian beam, the PSD is proportional to the degree of ellipticity of the polarization. In our experiment, we control the ellipticity of the target beam using a linear polarizer (LP) and  quarter-wave plate (QWP). Fig. 3C shows the observed dependence of the effective field on the angle of the QWP ($\theta$). Also plotted are a sinusoidal fit of the data (red dashed curve) and the result of full wave numerical simulations (purple curve, see supplementary information). The data shows a small DC offset in the curve which is not present in Eq. 1. This offset is related to an asymmetry in the system's geometry, which would lead to the presence of transverse spin at the location of the NV center. This transverse spin obeys spin-momentum locking rules  \cite{mechelen_universal_2016,kalhor_universal_2016} and does not depend on the QWP angle and therefore, gives rise to the offset. This non-ideality is captured by our simulations (see supplementary information D).

The second convincing proof of PSD measurement is the linear dependence of the effective field on the power of the beam (Eq. 1) shown in Fig. 3D. The PSD is linearly proportional to the power of the beam. This is in contrast to real magnetic fields where the amplitude scales with the square root of the power. The dashed red line shows a linear fit to the measured data. 
It should be noted that for each data point in Fig. 3C\&D we measure the energy shifts $\delta_{\pm 1}-\delta_0$ separately and calculate the effective field $B_{eff}$ according to Eq. 1 (see supplementary information C). The inset of these figures show the raw data for these measurements.

\begin{figure}
    \centering
    \includegraphics[width=160mm]{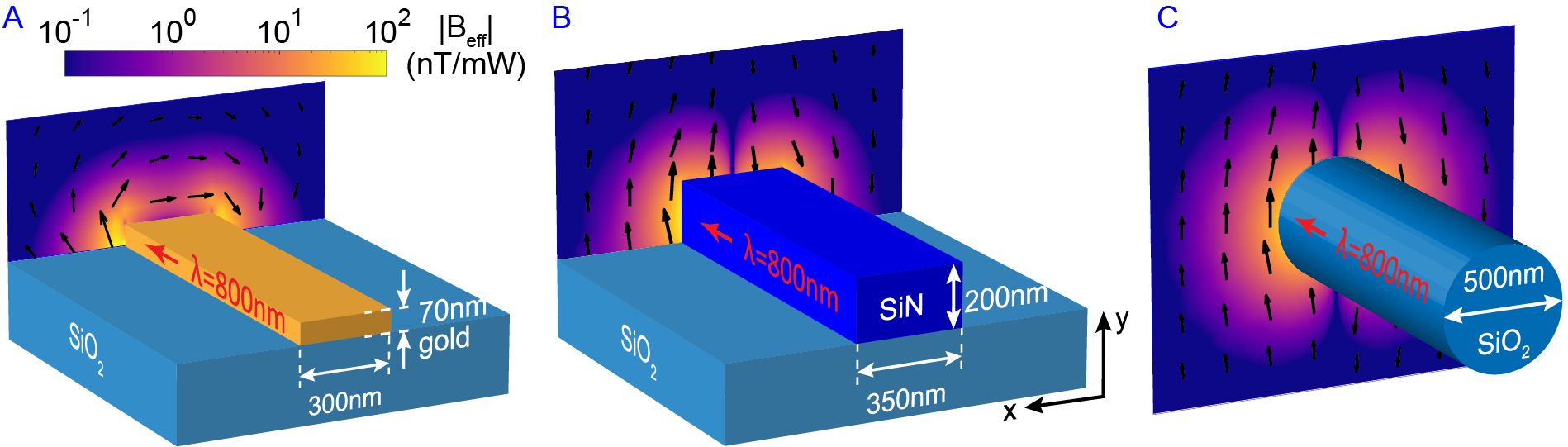}
    \caption{Effective field sensed by an NV center near optical waveguides. A plasmonic waveguide (A), a ridge waveguide (B), and an optical fiber (C) all producing transverse effective fields due to the PSD of their evanescent fields. The direction of the effective field is shown with black arrows and its amplitude with the colormap. The mode of the ridge waveguide is transverse electric (TE) and the mode of the optical fiber is $\hat{x}$ polarized $HE_{11}$.}
    \label{fig4}
\end{figure}

We now discuss how this universal photon-spin-density induced effective static magnetic fields can usher a new generation of on-chip spin QED applications. We note that evanescent waves are a ubiquitous resource available on a scalable nanophotonic platform. These evanescent waves possess an intrinsic universal spin that can exert ultra-subwavelength resolution static effective fields on spin qubits. This effective magnetic field is only manifested on interaction of PSD with NV centers and is a synthetic magnetic field. PSD induced magnetic fields can generate effective fields with giant spatial gradients on the order of  $10T/m$ with an ultrafast temporal response \cite{mitsch_exploiting_2014}. This allows for on-chip and targeted nanoscale addressing of spin qubits. Fig. 4A\&B\&C show the effective magnetic field in the near-field region of a plasmonic waveguide, a ridge waveguide, and an optical fiber. This static magnetic field  originates solely from the PSD of the evanescent waves in the vicinity of the waveguides. The direction of the field is in $x-y$ plane and is shown with black arrows. All three cases show short range effective magnetic field suitable for addressing single NV centers on chip or deposited on the surface of an optical fiber.

In this paper, we have demonstrated a room temperature quantum probe for nanoscale spinning light fields. The ultra-subwavelength behavior of spin angular momentum possess unique challenges for direct observation which we overcome using state-of-the-art quantum sensing approaches. Our measurement reveals that the NV center's room temperature excited state energy levels display striking agreement with those found in low temperature measurements. This is in contrast to direct measurements on the excited state where its full features cannot be revealed due to time averaging \cite{neumann_excited-state_2009,rogers_time_2009}. Our work can lead to new spin-dependent topological phases of light and also presents a way to exploit the universal resource of spinning evanescent waves available on an integrated photonics platform.

\section*{Acknowledgements}
Authors acknowledge funding from DARPA Nascent Light-Matter Interactions. Leif Bauer acknowledges the National Science Foundation for support under the Graduate Research Fellowship Program (GRFP) under grant number DGE-1842166.

\baselineskip0pt
\bibliography{Bib_FF}
\bibliographystyle{Science}

\end{document}

\baselineskip14pt

% Make the title.

\maketitle 

\renewcommand{\figurename}{Fig.}
\renewcommand{\thefigure}{S\arabic{figure}}
\setcounter{figure}{0}
\renewcommand\thesection{\Alph{section}}

%%%%%%%%%%%%%%%%%%%%%%%%%%%%%%%%%%%%%%%%%%%%%%%%%%%%%%%%%%%%%%
\section{Virtual transition induced effective magnetic field}
We now show how to calculate the effective field induced by a far off-resonant electric dipole transitions. Virtual electric dipole transitions will induce 
energy shifts in the ground states, which can be described by an effective Hamiltonian $H_{\rm shift} = \sum_i \delta_i |i\rangle\langle i|$ with the shift ({\it 37, 38\/})
\begin{equation}
\delta_i =\frac{1}{\hbar^2} \sum_{f} \frac{ \left| \langle f|\boldsymbol{d}|i\rangle\cdot \boldsymbol{E}^{+}(\boldsymbol{r},t) \right|^2\ } {\Delta_{if}+\Gamma_{f}^2/4\Delta_{if}}.
\end{equation}
Here, $|i\rangle$ and $|f\rangle$ are the initial and final states of the possible transitions in an NV center as shown in Fig.~\ref{fig:S1}, $\Delta_{ij}$ is the difference between the center frequency of the off-resonant excitation ($\omega_0$) and the resonance frequency of the transition ($\omega_{ij}$), $\Gamma_f$ is the spontaneous decay rate of the final state, and $\boldsymbol{E}^{+}(\boldsymbol{r},t)$ is the positive frequency part of the electric field of the excitation at the position of the NV ($\boldsymbol{r}$). Usually, the detuning is much larger than the spontaneous decay rate. Thus, the $\Gamma_f^2$ term will be neglected. We show that the energy difference in the states $\{|0\rangle,|\pm1\rangle\}$ induced by the off-resonance light functions as an effective static magnetic field for the NV ground-state spin.

\subsection{Energy structure of NV center}
To calculate this effective field, we first give the eigenenergy spectrum and the possible transitions in the NV center. We only consider the six triplet excited states and neglect the other singlet states ({\it 39, 40\/}), as the electric dipole transitions do not change spin states. For convenience, we choose a basis set, which is both spin- and orbital angular momentum resolved:
\begin{align}
\left|E_{L}\right\rangle  & =\frac{1}{\sqrt{2}}\left(\left|E_{y}\right\rangle -i\left|E_{x}\right\rangle \right) =\frac{i}{2}\left[\left|a_{1}\right\rangle\left|e_{+}\right\rangle -\left|e_{+}\right\rangle\left|a_{1}\right\rangle \right]\otimes\left(\left|\uparrow\downarrow\rangle+|\downarrow\uparrow\right\rangle \right),\\
\left|E_{R}\right\rangle  & =\frac{1}{\sqrt{2}}\left(\left|E_{y}\right\rangle +i\left|E_{x}\right\rangle \right) =\frac{i}{2}\left[\left|a_{1}\right\rangle \left|e_{-}\right\rangle-\left|e_{-}\right\rangle \left|a_{1}\right\rangle \right]\otimes\left(\left|\uparrow\downarrow\rangle+|\downarrow\uparrow\right\rangle \right),
\end{align}
\begin{align}
\left|E_{\downarrow}\right\rangle  & = \frac{1}{\sqrt{2}}\left(\left|E_{2}\right\rangle +\left|E_{1}\right\rangle \right) =   \frac{1}{\sqrt{2}}\left(\left|a_{1}\right\rangle \left|e_{-}\right\rangle - \left|e_{-}\right\rangle \left|a_{1}\right\rangle \right)\otimes\left|\downarrow\downarrow\right\rangle  ,\\
\left|E_{\uparrow}\right\rangle  & =\frac{1}{\sqrt{2}}\left(\left|E_{2}\right\rangle -\left|E_{1}\right\rangle \right)=\frac{1}{\sqrt{2}}\left(\left|a_{1}\right\rangle \left|e_{+}\right\rangle -\left|e_{+}\right\rangle \left|a_{1}\right\rangle \right)\otimes\left|\uparrow\uparrow\right\rangle ,
\end{align}
\begin{align}
\left|A_{\uparrow}\right\rangle  & =\frac{1}{\sqrt{2}}\left(\left|A_{2}\right\rangle +\left|A_{1}\right\rangle \right)=\frac{1}{\sqrt{2}}\left(\left|a_{1}\right\rangle \left|e_{-}\right\rangle -\left|e_{-}\right\rangle \left|a_{1}\right\rangle \right)\otimes\left|\uparrow\uparrow\right\rangle ,\\
\left|A_{\downarrow}\right\rangle  & =\frac{1}{\sqrt{2}}\left(\left|A_{2}\right\rangle -\left|A_{1}\right\rangle \right)=\frac{1}{\sqrt{2}}\left(\left|a_{1}\right\rangle \left|e_{+}\right\rangle -\left|e_{+}\right\rangle \left|a_{1}\right\rangle \right)\otimes\left|\downarrow\downarrow\right\rangle ,
\end{align}
where $|a_1\rangle$, $|e_x\rangle$, and $|e_y\rangle$ are the orbital states of the NV center and $\left|e_{\pm}\right\rangle =\mp\left(\left|e_{x}\right\rangle \pm i\left|e_{y}\right\rangle \right)/\sqrt{2}$.

\begin{figure}[bt]
    \centering
    \includegraphics[width=16cm]{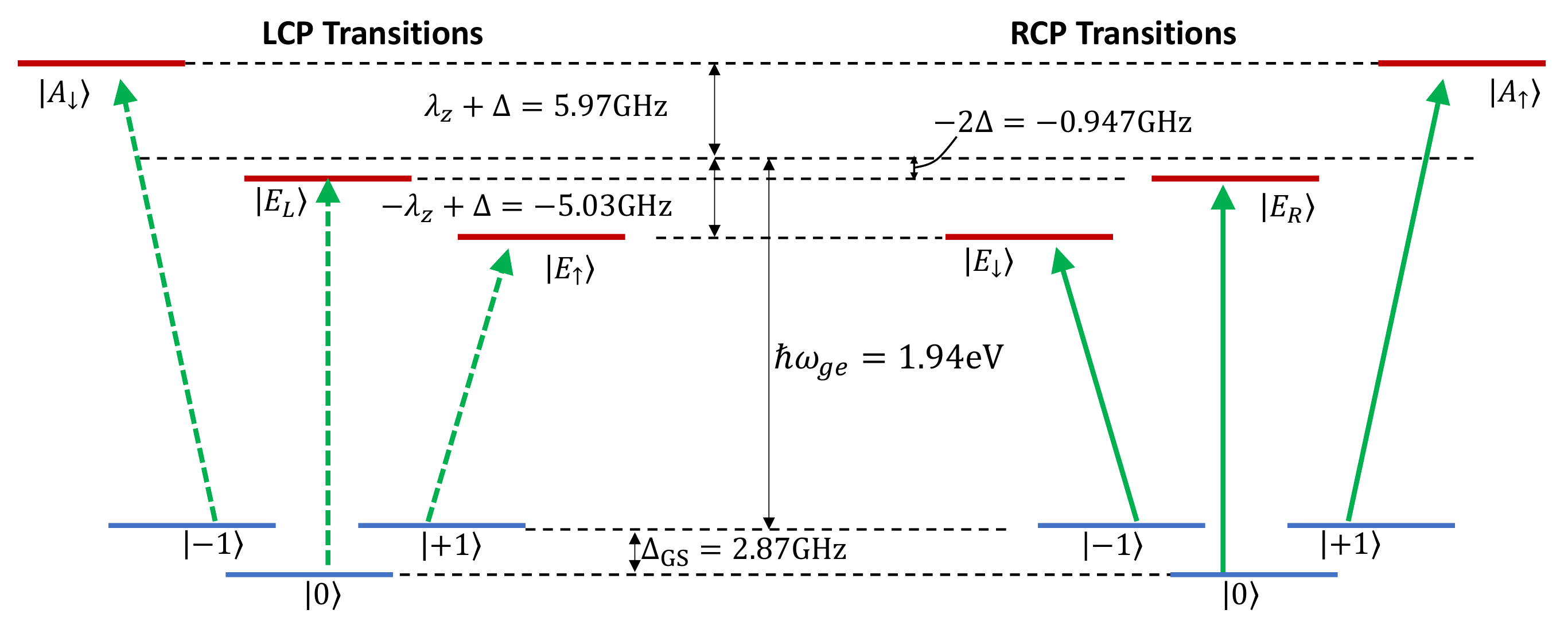}
    \caption{\label{fig:S1} Schematic of energy levels of the NV center. Here, there are no external magnetic field and strain. Left and right panels show the possible transitions induced by left circularly polarized (LCP) and right circularly polarized (RCP) lights.}
\end{figure}

The ground states are given by,
\begin{align}
\left|-1\right\rangle  & =\frac{1}{\sqrt{2}}\left(\left|e_{x}e_{y}\right\rangle -\left|e_{y}e_{x}\right\rangle \right)\otimes\left|\downarrow\downarrow\right\rangle ,\\
\left|0\right\rangle  & =\frac{1}{2}\left(\left|e_{x}e_{y}\right\rangle -\left|e_{y}e_{x}\right\rangle \right)\otimes\left(\left|\uparrow\downarrow\right\rangle +\left|\downarrow\uparrow\right\rangle \right),\\
\left|+1\right\rangle  & =\frac{1}{\sqrt{2}}\left(\left|e_{x}e_{y}\right\rangle -\left|e_{y}e_{x}\right\rangle \right)\otimes\left|\uparrow\uparrow\right\rangle .
\end{align}
We note that this normalization is different from Ref. ({\it 39, 40\/}). The possible electric dipole transition are shown in Fig.~\ref{fig:S1}. 

\subsection{NV Center Hamiltonian}
The effective excited-state Hamiltonian of the NV center is given by,
\begin{align}
H_{\rm ES} = &  (\gamma_{{\rm NV}}B+\Delta_{es}+\lambda_{z})|A_{\uparrow}\rangle\langle A_{\uparrow}| +(-\gamma_{{\rm NV}}B+\Delta_{es}+\lambda_{z})|A_{\downarrow}\rangle\langle A_{\downarrow}| - 2\Delta_{es} |E_{R}\rangle\langle E_{R}|  \nonumber\\
& - 2\Delta_{es} |E_{L}\rangle\langle E_{L}|+(\gamma_{{\rm NV}}B+\Delta_{es}-\lambda_{z})|E_{\uparrow}\rangle\langle E_{\uparrow}|+(-\gamma_{{\rm NV}}B+\Delta_{es}-\lambda_{z})|E_{\downarrow}\rangle\langle E_{\downarrow}|,
\end{align}
where $\lambda_z\approx 2\pi\times 5.5$~GHz is the spin-orbit coupling, $\Delta_{es} \approx2\pi\times 1.42/3$~GHz is the spin-spin induced zero-field splitting, the optical gap $\hbar\omega_{ge}\approx 1.945$~eV ($637$~nm) between the ground states and the excited states (see Fig.~\ref{fig:S1}) has not been shown in $H_{\rm ES}$. The off-diagonal coupling between these six excited states does not change this virtual transition induced effective magnetic. Thus, we have omitted those off-diagonal coupling terms and the excited state Hamiltonian is of diagonal form $H_{\rm ES}=E_j |j\rangle\langle j|$ where $|j\rangle \in \{\left|A_{\uparrow}\right\rangle ,\left|A_{\downarrow}\right\rangle ,\left|E_{R}\right\rangle ,\left|E_{L}\right\rangle ,\left|E_{\uparrow}\right\rangle ,\left|E_{\downarrow}\right\rangle \}$.

\subsection{Energy shifts and effective magnetic field for the ground-state spin}
We now also give the transition elements of the electric dipole transitions in the NV center. As the upper basis is based on the two-hole picture, the electric dipole operator of the NV is given by
\begin{equation}
\boldsymbol{d}=e(\boldsymbol{r}_{1}+\boldsymbol{r}_{2}),
\end{equation}
where $\boldsymbol{r}_{1}$ and $\boldsymbol{r}_{2}$ are the position operator of the two holes.
According to the symmetry of the orbits shown in ({\it 40\/}),
the nonzero elements of the dipolar transitions are
\begin{equation}
e\left\langle e_{x}\right|r_{x}\left|e_{x}\right\rangle =-e\left\langle e_{y}\right|r_{y}\left|e_{y}\right\rangle =e\left\langle e_{y}\right|r_{y}\left|e_{x}\right\rangle \neq0.
\end{equation}
and 
\begin{equation}
e\left\langle a_{1}\right|r_{x}\left|e_{x}\right\rangle =e\left\langle a_{1}\right|r_{y}\left|e_{y}\right\rangle \equiv d_{0}\neq0.
\end{equation}

\begin{table}
\centering
\begin{tabular}{|c|c|c|c|c|c|c|}
\hline 
Polarization & $\left|A_{\uparrow}\right\rangle $ & $\left|A_{\downarrow}\right\rangle $ & $\left|E_{R}\right\rangle $ & $\left|E_{L}\right\rangle $ & $\left|E_{\uparrow}\right\rangle $ & $\left|E_{\downarrow}\right\rangle $\tabularnewline
\hline 
\hline 
$\left|-1\right\rangle $ &  & LCP &  &  &  & RCP\tabularnewline
\hline 
$\left|0\right\rangle $ &  &  & RCP & LCP &  & \tabularnewline
\hline 
$\left|+1\right\rangle $ & RCP &  &  &  & LCP & \tabularnewline
\hline 
\end{tabular}
\caption{\label{tab:Selection}Selection rules for optical transitions between
the triplet ground states and the triplet excited states. We note
that any transition connected with circularly polarized light can
also be stimulated with linearly polarized lights. But the corresponding transition strength will be smaller.}
\end{table}

The possible transitions and the corresponding transition strength can be easily obtained by the transition element. The non-zero transition elements are: 
\begin{align}
\left\langle E_{L}\right|\boldsymbol{d}\left|0\right\rangle  & = d_{0}\boldsymbol{e}_{R},\ \left\langle E_{R}\right|\boldsymbol{d}\left|0\right\rangle = d_{0}\boldsymbol{e}_{L},\\
\left\langle A_{\uparrow}\right|\boldsymbol{d}\left|+1\right\rangle  & = id_{0}\boldsymbol{e}_{L},\ \left\langle E_{\uparrow}\right|\boldsymbol{d}\left|+1\right\rangle = id_{0}\boldsymbol{e}_{R}\\
\left\langle A_{\downarrow}\right|\boldsymbol{d}\left|-1\right\rangle  & = id_{0}\boldsymbol{e}_{R},\ \left\langle E_{\downarrow}\right|\boldsymbol{d}\left|-1\right\rangle = id_{0}\boldsymbol{e}_{L},
\end{align}
where we have defined the unit vectors
\begin{equation}
\boldsymbol{e}_{L} =(\boldsymbol{e}_{x}+i\boldsymbol{e}_{y})/\sqrt{2},\ \boldsymbol{e}_{R}  =(\boldsymbol{e}_{x}-i\boldsymbol{e}_{y})/\sqrt{2}.    
\end{equation}
Using the identities of the unit vectors $\boldsymbol{e}_{R}\cdot\boldsymbol{e}^{*}_{R}=\boldsymbol{e}_{R}\cdot\boldsymbol{e}_{L}=1$
and $\boldsymbol{e}_{R}\cdot\boldsymbol{e}_{R}=\boldsymbol{e}_{L}\cdot\boldsymbol{e}_{L}=0$,
one can easily obtain the selection rules in Table~\ref{tab:Selection}.

The value of the transition element can be obtained from the life times of the triplet excited states:
\begin{equation}
\tau_{\rm NV}=\frac{1}{\gamma_{\rm NV}}\approx\left(\frac{\omega_{eg}^{3}d_{0}^{2}}{3\pi\hbar\varepsilon_{0}c^{3}}\right)^{-1}.
\end{equation}
In this work, the lifetime of the NV center is taken as $\tau_{\rm NV}=15$~ns.
Then, we have $d_0\approx 2.485\times 10^{-29}\, {\rm C\cdot m}$.

For simplicity, we first consider the case where the NV center axis is aligned with the propagating direction of the off-resonant excitation ($+\hat{z}$). The energy shifts in the ground states sub-levels under a LCP or RCP excitation field are given by
\begin{align}
\delta_{-1,L} & =\frac{1}{2\hbar^2}\frac{d_{0}^{2}\Delta_{-1,A_{\downarrow}}}{\Delta_{-1,A_{\downarrow}}^{2}+\Gamma^{2}/4}|\boldsymbol{E}(\boldsymbol{r})|^2,\ \delta_{-1,R}=\frac{1}{2\hbar^2} \frac{d_{0}^{2}\Delta_{-1,E_{\downarrow}}}{\Delta_{-1,E_{\downarrow}}^{2}+\Gamma^{2}/4}|\boldsymbol{E}(\boldsymbol{r})|^2, \label{eq:shift_m1} \\
\delta_{0,L} & =\frac{1}{2\hbar^2}\frac{d_{0}^{2}\Delta_{0,E_{L}}}{\Delta_{0,E_{L}}^{2}+\Gamma^{2}/4}|\boldsymbol{E}(\boldsymbol{r})|^2,\ \delta_{0,R} =\frac{1}{2\hbar^2}\frac{d_{0}^{2}\Delta_{0,E_{R}}}{\Delta_{0,E_{R}}^{2}+\Gamma^{2}/4}|\boldsymbol{E}(\boldsymbol{r})|^2 ,\label{eq:shift_0}\\
\delta_{+1,L} & =\frac{1}{2\hbar^2}\frac{d_{0}^{2}\Delta_{+1,E_{\uparrow}}}{\Delta_{+1,E_{\uparrow}}^{2}+\Gamma^{2}/4}|\boldsymbol{E}(\boldsymbol{r})|^2,\ \delta_{+1,R}=\frac{1}{2\hbar^2} \frac{d_{0}^{2}\Delta_{+1,A_{\uparrow}}}{\Delta_{+1,A_{\uparrow}}^{2}+\Gamma^{2}/4}|\boldsymbol{E}(\boldsymbol{r})|^2,\label{eq:shift_1}
\end{align}
where we have used the relation $|\boldsymbol{E}(\boldsymbol{r})|^2\approx 2\boldsymbol{E}^{-}(\boldsymbol{r})\cdot\boldsymbol{E}^{+}(\boldsymbol{r})$ and the detunings are given by
\begin{align}
\Delta_{-1,j} & =\omega_{0}-(\omega_{ge}+E_{j}+\gamma_{{\rm NV}}B),\\
\Delta_{0,j} & =\omega_{0}-(\omega_{ge}+E_{j}+\Delta_{{\rm GS}}),\\
\Delta_{+1,j} & =\omega_{0}-(\omega_{ge}+E_{j}-\gamma_{{\rm NV}}B).
\end{align}

\subsection{Effective magnetic field for probe qubits}
In the experiment, we choose two ground states sub-levels to form a qubit to detect the relative energy shift between them. The effective magnetic fields for the three possible qubits are defined as
\begin{align}
B_{01,L}& =\frac{\delta_{+1,L}-\delta_{0,L}}{\gamma_{NV}},\ B_{01,R} =\frac{\delta_{+1,R}-\delta_{0,R}}{\gamma_{NV}}, \\
B_{-10,L} & =\frac{\delta_{0,L}-\delta_{-1,L}}{\gamma_{NV}},\ B_{ -10,R} =\frac{\delta_{0,R}-\delta_{-1,R}}{\gamma_{NV}} \\
B_{-11,L} & =\frac{\delta_{+1,L}-\delta_{-1,L}}{2\gamma_{NV}},\ B_{-11,R} =\frac{\delta_{+1,R}-\delta_{-1,R}}{2\gamma_{NV}}.
\end{align}

\begin{figure}[tb]
\centering
\includegraphics[width=12cm]{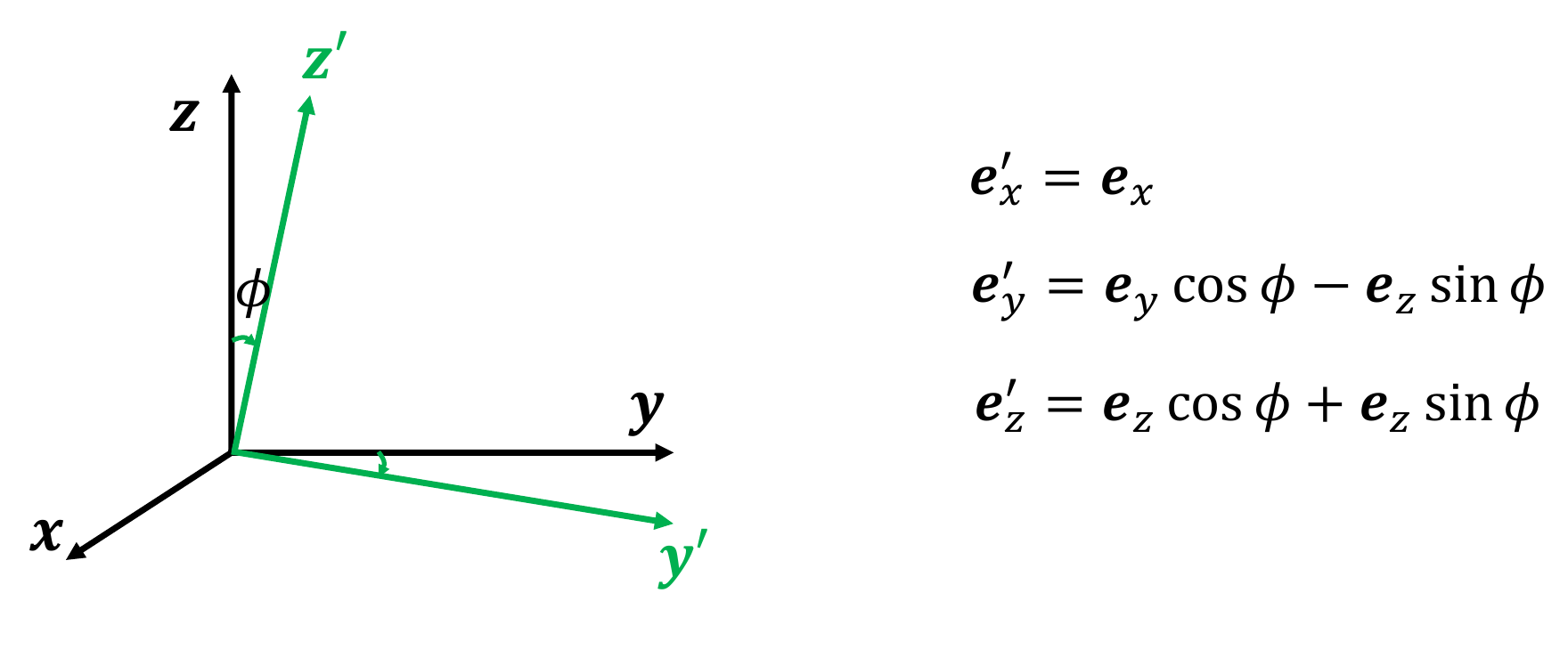}
\caption{\label{fig:S2} Transformation between the NV center coordinate frame and the off-resonant beam coordinate frame.}
\end{figure}

We note that the magnitude of the effective field is linearly proportional to the off-resonant laser power. We can also tune the effective field via changing the polarization of the laser. The polarization unit vector of an arbitrary polarized light can be expanded as 
\begin{equation}
\boldsymbol{e}=\boldsymbol{e}_L\cos\left(\theta-\frac{\pi}{4}\right)+\boldsymbol{e}_R\sin\left(\theta-\frac{\pi}{4}\right),  
\end{equation}
where the angle $\theta_{\rm}\in [0,2\pi)$ is the rotation angle of a quarter-wave plate (QWP). In addition, in the experiment the NV center axis makes an angle $\phi$ with the propagating direction of the off-resonant beam. We need to transform the LCP and RCP unit vectors in the NV center coordinate frame ($x'y'z'$) into the excitation beam coordinate frame ($xyz$) as shown in Fig.~\ref{fig:S2},
\begin{align}
\boldsymbol{e}'_L & = \frac{1}{2}\boldsymbol{e}_L(\cos\phi+1) +\frac{1}{2}\boldsymbol{e}_R (\cos\phi -1)+ \frac{1}{\sqrt{2}}\boldsymbol{e}_z\sin\phi, \\
\boldsymbol{e}'_R & = \frac{1}{2}\boldsymbol{e}_R(\cos\phi+1) +\frac{1}{2}\boldsymbol{e}_L (\cos\phi -1)+ \frac{1}{\sqrt{2}}\boldsymbol{e}_z\sin\phi.
\end{align}
In this case, the three effective fields are given by
\begin{align}
B_{01} & = B_{01,L}|\boldsymbol{e}\cdot\boldsymbol{e}^{\prime *}_L|^2 + B_{01,R}|\boldsymbol{e}\cdot\boldsymbol{e}^{\prime *}_R|^2,\\
B_{-10} & = B_{-10,L}|\boldsymbol{e}\cdot\boldsymbol{e}^{\prime *}_L|^2 + B_{-10,R}|\boldsymbol{e}\cdot\boldsymbol{e}^{\prime *}_R|^2,\\
B_{-11} &= B_{-11,L}|\boldsymbol{e}\cdot\boldsymbol{e}^{\prime *}_L|^2 + B_{-11,R}|\boldsymbol{e}\cdot\boldsymbol{e}^{\prime *}_R|^2.
\end{align}
Using the fact that $\delta_{-1,L}=\delta_{+1,R}$ and $\delta_{-1,R}=\delta_{+1,L}$, we can rewrite $B_{-11}$ as
\begin{align}
B_{-11} & = \frac{\delta_{+1,L}-\delta_{+1,R}}{2\gamma_{NV} |\boldsymbol{E}(\boldsymbol{r})|^2}\{-i[\boldsymbol{E}^{*}(\boldsymbol{r})\times\boldsymbol{E}(\boldsymbol{r})]\cdot\boldsymbol{\hat{n}}\}\\
& = \frac{2\omega_0(\delta_{+1,L}-\delta_{+1,R})}{\epsilon\gamma_{NV} |\boldsymbol{E}(\boldsymbol{r})|^2}\{\boldsymbol{S}_E\cdot\boldsymbol{\hat{n}}\}\\
& \approx -\frac{2\omega_0 d^2_0\lambda_z}{\hbar^2\epsilon\gamma_{NV} \Delta_{+1,A_{\uparrow}}\Delta_{+1,E_{\uparrow}}}\{\boldsymbol{S}_E\cdot\boldsymbol{\hat{n}}\}.
\end{align}
Here, $\epsilon$ is the permittivity of diamond, $\boldsymbol{S}_E=-(i\epsilon/4\omega)\boldsymbol{E}^*\times\boldsymbol{E}$ is the electric contribution to photonic spin density, and $\hat{n}$ is the direction of the NV center. We have also used the relation,
\begin{equation}
\boldsymbol{e}^{*}\times\boldsymbol{e}=i\left[\cos^2 \left(\theta-\frac{\pi}{4}\right)-\sin^2 \left(\theta-\frac{\pi}{4}\right)\right]\boldsymbol{e}_z=i \, sin(2\theta)\boldsymbol{e}_z. 
\end{equation}

In the far off-resonant case $\Delta_{+1,A_{\uparrow}}\approx\Delta_{+1,E_{\uparrow}}=\Delta$. In this case the denominator is only determined by the frequency of the off-resonant beam and the spin-orbit coupling factor in the excited state. $B_{eff}$ can be written in the following form,
\begin{align}
B_{eff}=B_{-11} & = -\frac{2\omega_0 d^2_0\lambda_z}{\hbar^2\epsilon\gamma_{NV} \Delta^2}(\boldsymbol{S}_E\cdot\boldsymbol{\hat{n}})=\frac{C}{\lambda_0\Delta^2}(\boldsymbol{S}_E\cdot\boldsymbol{\hat{n}})
\end{align}
where $C$ is a constant and $\lambda_0$ is the wavelength of the off-resonant excitation.

%%%%%%%%%%%%%%%%%%%%%%%%%%%%%%%%%%%%%%%%%%%%%%%%%%%%%%%%%%%%%%
\section{Experimental setup}

\begin{figure}[bt]
    \centering
    \includegraphics[width=120mm]{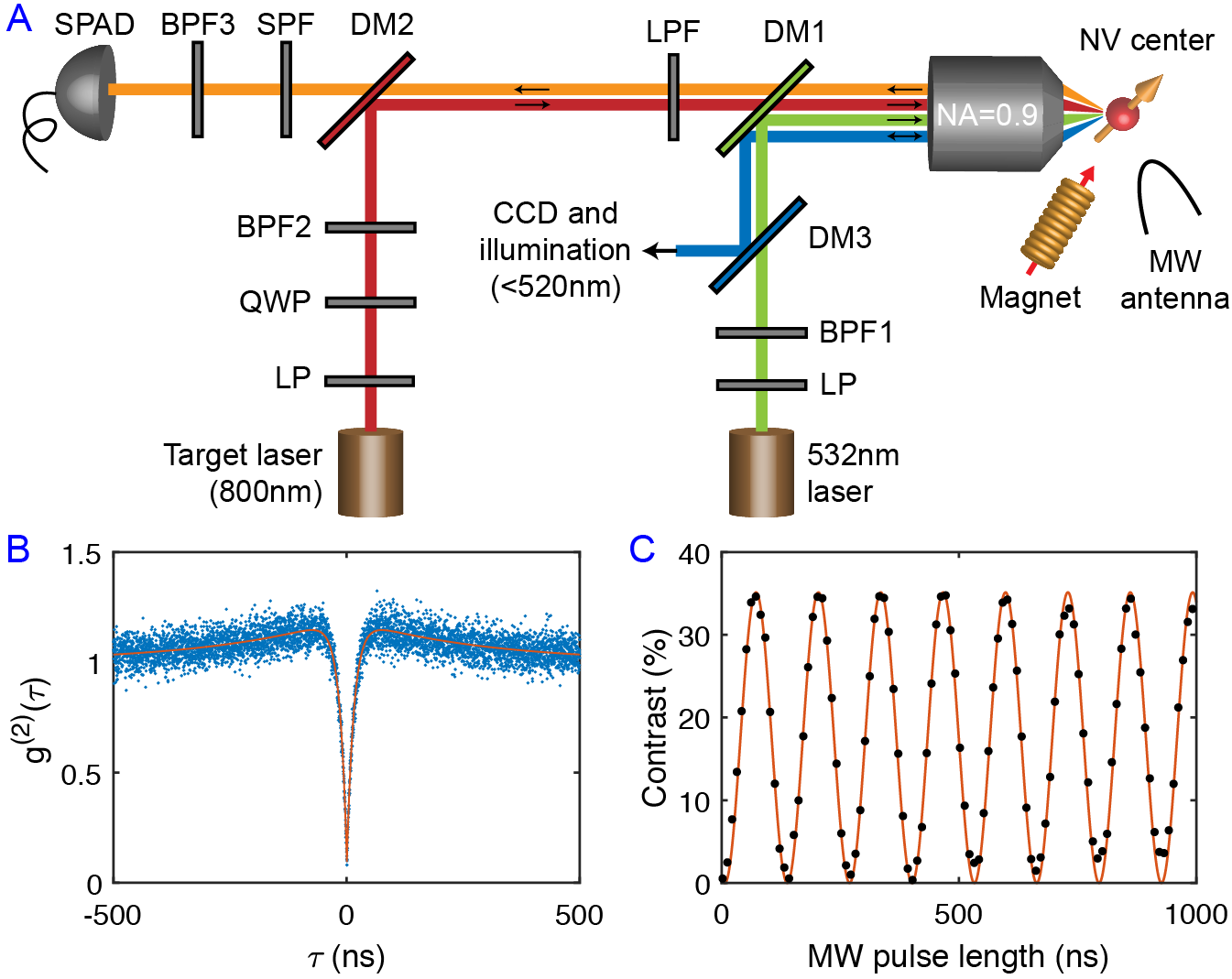}
    \caption{Schematic of the experimental setup. (A) Schematic of the setup showing the beam paths and filters. (B) Second-order correlation measurement showing that the AFM tip contains a single NV center. (C) Rabi oscillations showing the interaction of the microwave beam (MW) with the NV center. LP, linear polarizer; DM, dichroic mirror; BPF, band-pass filter; LPF, long-pass filter; SPF, short-pass filter; QWP, quarter-wave plate.}
    \label{fig:exp_setup}
\end{figure}

A schematic of the experimental setup is shown in Fig. \ref{fig:exp_setup}A. The setup comprises an NV center and three main beam paths. The NV center (Fig. \ref{fig:exp_setup}B) is on an AFM tip and is implanted $10nm$ deep into the surface of the tip. The tip is purchased from QZabre LLC. The second-order correlation measurement for the single NV center is shown in Fig. \ref{fig:exp_setup}B. A magnet breaks the degeneracy of $|\pm 1 \rangle$ states by applying a magnetic field $B_{bias}\approx 1.1 mT$. An antenna made of a $15\mu m$ thick tungsten wire delivers a MW signal to the NV center to induce Rabi oscillations (Fig. \ref{fig:exp_setup}C). 

A $532nm$ laser is used for excitation and readout of the state of the NV center. The beam is chopped with an AOM for pulsed measurements. After the linear polarizer the beam is coupled to a polarization maintaining (PM) fiber. The output of the fiber is collimated and filtered with a band-pass filter (BPF) (Thorlabs FLH532-10).

The target laser is a TTL controlled laser diode module (Power Technology Inc. PMT150). The beam is coupled into a PM fiber after a linear polarizer. The output of the fiber is collimated and goes through a quarter-wave plate before being filtered with a BPF (FBH800-10). Due to reflection from a dichroic mirror, the polarization of the beam has an ellipticity of $\varepsilon=1.08$ before entering the objective lens when the QWP is at its optimal angle.

The photoluminescence signal from the NV center is filtered with a long-pass filter (Semrock BLP02-561-R), a short-pass filter (Semrock SP01-785RU), and a BPF (Semrock FF01-709/167). Three dichroic mirrors (DM) combine and separate the beams; DM1, Thorlabs DMLP550R; DM2, Semrock FF750-SDi02; DM3, Semrock FF520-Di02. The PL signal is detected by a Micro Photon Devices SPAD. For measuring $g^{(2)}$ the signal is coupled to a fiber optic beam splitter and is detected with two SPADs.

%%%%%%%%%%%%%%%%%%%%%%%%%%%%%%%%%%%%%%%%%%%%%%%%%%%%%%%%%%%
\section{Pulse sequence}

\begin{figure}[bt]
    \centering
    \includegraphics[width=160mm]{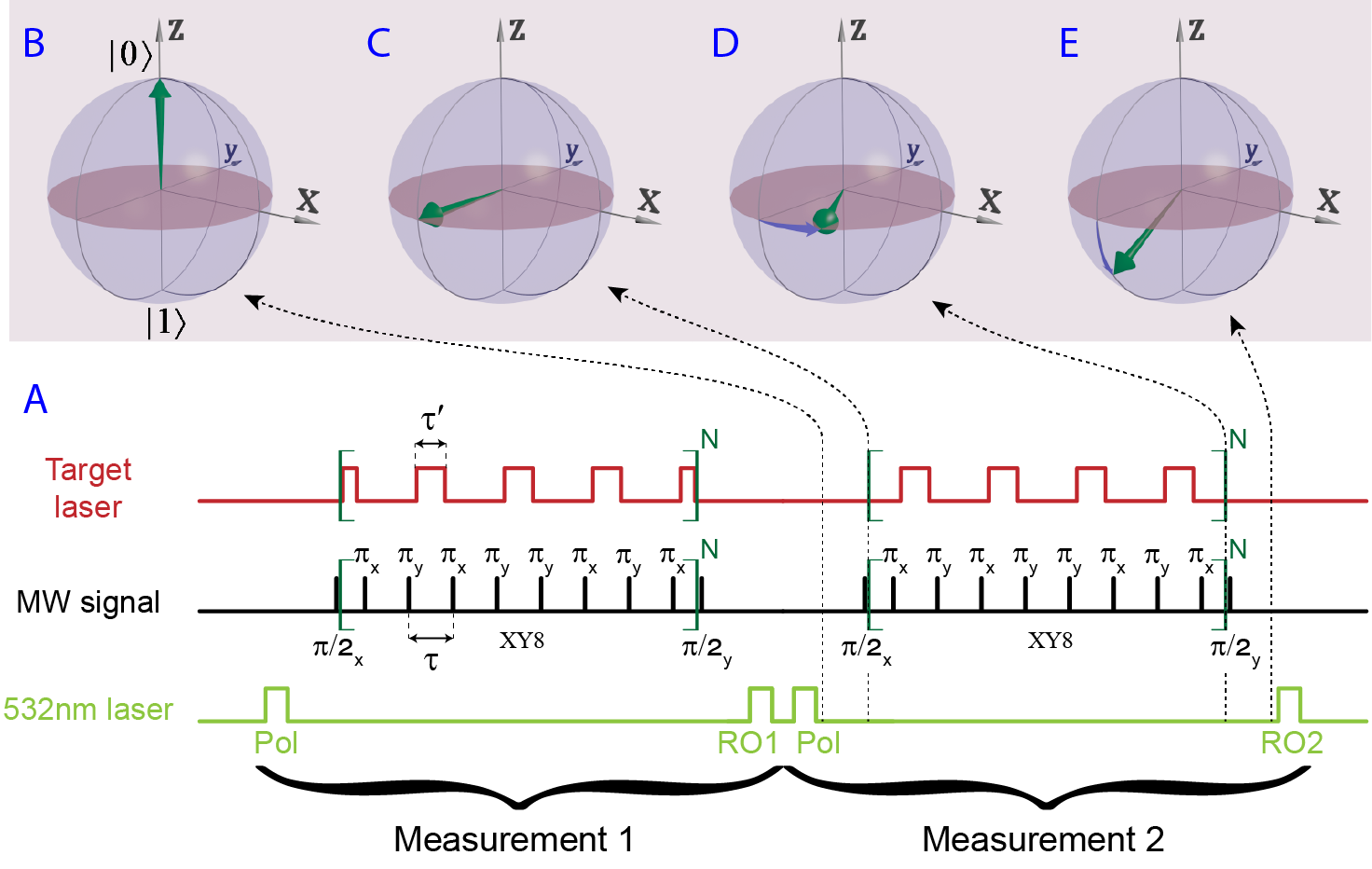}
    \caption{(A) Pulse sequence used for measuring the PSD. In each measurement a pair of 532nm laser pulses initialize (polarizing) and readout the state of the NV center. During each measurement an XY8 MW pulse performs dynamical decoupling on the spin of the NV center. (B-E) Bloch sphere representation of the state of the qubit after polarization, before the XY8 pulse, after the XY8 pulse, and before readout. The PSD is generated by the target beam. RO, readout; Pol, polarizing.}
    \label{fig:pulse_sequence}
\end{figure}

Here we will explain how the measurement is performed in details. Depending on which transition the MW field is resonant to ($|0\rangle$ to $|+1\rangle$ or $|0\rangle$ to $|-1\rangle$) we choose those two states as a two-level system. The pulse sequence, shown in Fig. \ref{fig:pulse_sequence}, starts with initializing the NV center in $|0\rangle$ state. Then a series of $\pi-$ and $\pi/2$-pulses manipulate the state of the NV center to prepare the NV center for interaction and readout and to increase the coherence time. The state after each pulse is calculated using the following rotation matrices,
$$R_x(\theta)=
\begin{bmatrix}
cos\frac{\theta}{2} & -isin\frac{\theta}{2} \\
-isin\frac{\theta}{2} & cos\frac{\theta}{2}
\end{bmatrix}
$$
$$and \hspace{10pt} R_y(\theta)=
\begin{bmatrix}
cos\frac{\theta}{2} & -sin\frac{\theta}{2} \\
sin\frac{\theta}{2} & cos\frac{\theta}{2}
\end{bmatrix}
,$$
where $\theta=\pi/2$ ($\theta=\pi$) for $\pi/2$-pulse ($\pi$-pulse).

We show the calculations for $|0\rangle$ to $|+1\rangle$ transition here. The sequence starts with initializing the NV center into $|0\rangle$ state. Then a $\pi/2$-pulse rotates the spin into a superposition state and a series of $\pi$ pulses perform the dynamical decoupling while the target beam affects the state of the NV center. The effect of each pulse of the target beam is captured by an added phase, $\phi$.
We are assuming that the MW pulses are very short and off-resonant to the NV center by $\omega-\omega_0=\delta\omega$.
This adds a phase $\Theta=\delta\omega\tau/2$ during each free procession period of length $\tau /2$. In our experiment we omit the first and last half target pulses in measurement 1 and omit the last target pulse in measurement 2 for simplicity. Therefore, each measurement contains total of $4N-1$ target pulses, where $N$ is the number of XY8 pulses.

After sending N nember of XY8 pulses, a total of 8N $\pi$-pulses have been sent. The state of the NV center before the last $\pi$/2-pulse is,
$$|\psi(N\tau)\rangle=
\frac{\sqrt{2}}{2}(e^{-i(4N-1)\phi}|0\rangle
-i e^{i(4N-1)\phi}|+1\rangle)
,$$

We set the last $\pi/2$ pulse in $-\hat{y}$ direction. The state of the NV center after this pulse is,
\begin{equation} \label{eq:final_state}
    |\psi(N\tau)\rangle=e^{-i\pi/4}(\cos(\pi/4-\Phi)|0\rangle
    -i\sin(\pi/4-\Phi)|+1\rangle),
\end{equation}
where $\Phi=(4N-1)\phi$ is the total phase induced by the target beam. Measured contrast for this state, $C_1$, is,
\begin{equation}
    C_1=C^{max}\left|\langle \pm 1 | \psi \rangle \right|^2=sin^2(\pi/4-\Phi)C^{max},
\end{equation}
where $C^{max}$ is the contrast measured for $|\pm 1\rangle$ states.

The outcome of Measurement 2 similarly will be,
\begin{equation}
    C_2=sin^2(\pi/4+\Phi)C^{max},
\end{equation}
where the difference in the sign of $\Phi$ is due to the target pulses being sent in the opposite time slots. We calculate the phase $\Phi$ by subtracting the outcome of the two measurements,
\begin{equation} \label{eq:contrast}
    C=C_1-C_2=\frac{C^{max}}{2}(cos(\pi/2+2\Phi)-cos(\pi/2-2\Phi))=-sin(2\Phi)C^{max}.
\end{equation}
In this approach, any undesired effect caused by adding the target beam (e.g., change of temperature and coherence) will be eliminated in the subtraction. 

To measure $C^{max}$, we employ the following procedure to ensure all the effects decreasing the coherence of the quantum states are accounted for. First we calculate the contrast $C^{ave}$ for the state  $|\psi\rangle=e^{-i\pi/4}(|0\rangle -i|+1\rangle)/\sqrt{2}$. Based on the measurements performed, $C^{ave}=(C_1+C_2)/2$. Then we measure the contrast $C^{+1}$ for the state $|\psi\rangle=-i|+1\rangle$. This measurement is performed with a pulse sequence similar to Measurements 1 and 2 with the difference that the last $\pi/2$ pulse is in $\hat{x}$ direction. Moreover, the target pulses are sent during both time slots to cancel out the effect of each other. The length of the pulses are set to halve of the pulses in Measurements 1 and 2 to keep the average power reaching the NV center same as measurements 1 and 2. This way any decoherence added to the system due to presence of the target beam is accounted for in measuring $C^{+1}$. Then we have,
\begin{equation} 
    C^{max}=2(C^{+1}-C^{ave}).
\end{equation}

If instead of the target pulses a magnetic field $B$ parallel to the axis of the NV center was applied, the measured contrast would have been,
\begin{equation} 
    C=sin(2\Phi)C^{max}=-sin((8N-2)\phi)C^{max},
\end{equation}
where $\phi=\gamma B \tau^{'} /2$. In this equation $\gamma$ is the gyromagnetic ratio for the NV center, and $\tau^{'}$ is the length of the target pulses. Therefore, we define an effective magnetic field equivalent to the effect caused by the target beam,
\begin{equation} 
\begin{split}
    B^{eff} &=\frac{-1}{(4N-1)\gamma \tau^{'}}sin^{-1}\left(\frac{C}{C^{max}}\right),\\
            &\approx \frac{-C}{(4N-1)\gamma \tau^{'} C^{max}}
\end{split}
\end{equation}

In our experiments, shown in Fig. 3 (main text), $\tau^{'}=1\mu s$ and N=4 is the number of XY8 pulses. For Fig. 3C $C^{max}$ is measured for the two measurements as, $C^{max}_{|0\rangle \rightarrow |-1\rangle}=20.3\% \pm 0.6$ and $C^{max}_{|0\rangle \rightarrow |+1\rangle}=19.2\% \pm 0.9$. For Fig. 3D $C^{max}$ is measured for the two measurements as, $C^{max}_{|0\rangle \rightarrow |-1\rangle}=20.2\% \pm 0.8$ and $C^{max}_{|0\rangle \rightarrow |+1\rangle}=19.1\% \pm 0.6$.

%%%%%%%%%%%%%%%%%%%%%%%%%%%%%%%%%%%%%%%%%%%%%%%%%%%%%%%%%%%
\section{Numerical simulations}

\begin{figure}
    \centering
    \includegraphics[width=\textwidth]{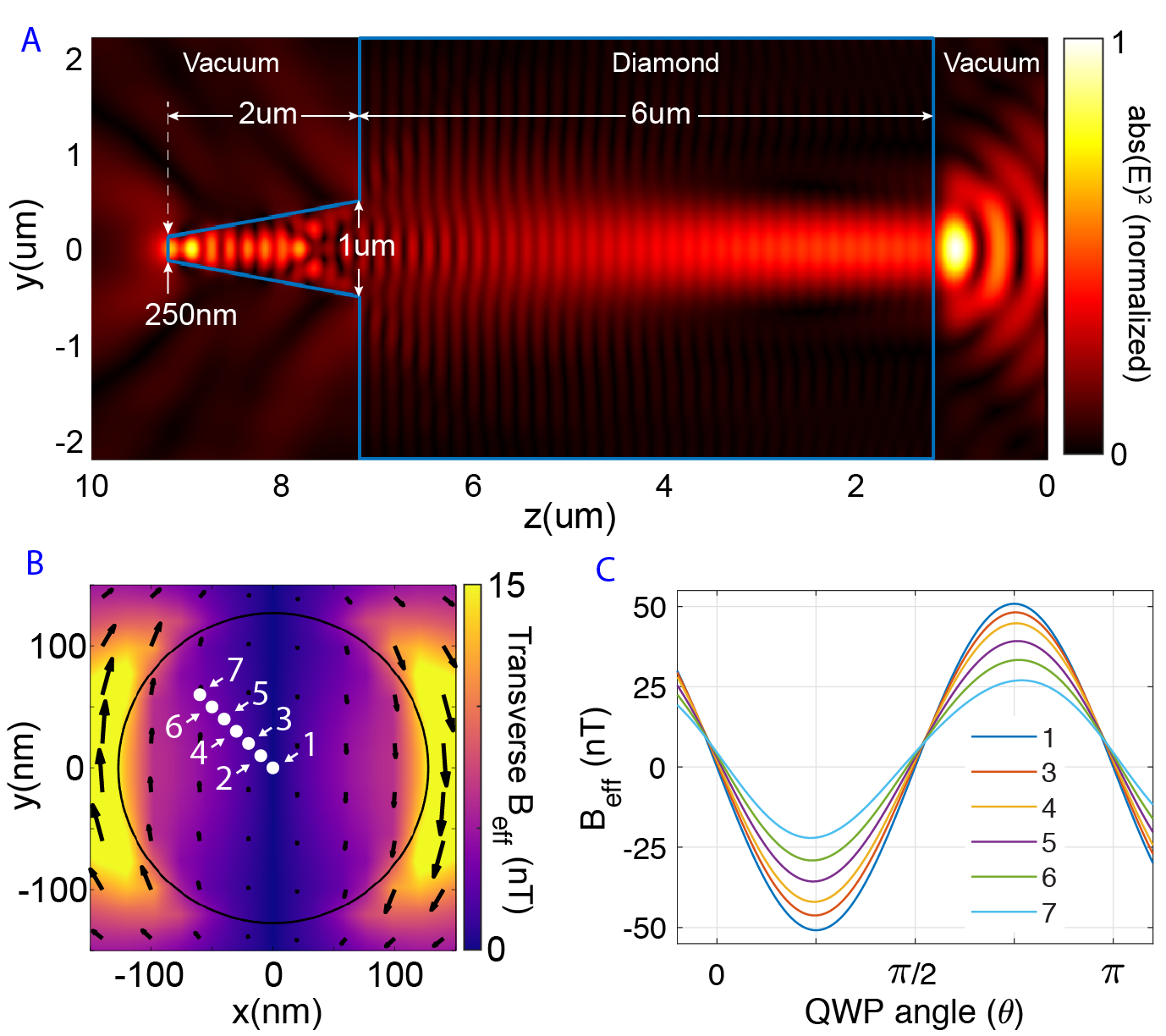}
    \caption{Numerical simulations for the effective field amplitude in an AFM tip. (A) Amplitude of the electric field in an AFM tip made of diamond. (B) Amplitude and direction of the transverse component of the effective field in the plane of the NV center. (C) The effective field seen by an NV center with direction $\hat{n}=(\hat{x}+\hat{y}-\hat{z})/\sqrt{3}$ placed at various locations shown in panel B with white dots.}
    \label{fig:simulations}
\end{figure}

Numerical simulations presented in Fig. 3C and Fig. 4 are performed with CST Studio Suite. The detail of the simulation results shown in Fig. 3C is explained here.  Fig.~\ref{fig:simulations}A shows the normalized amplitude of the electric field in the structure. The structure is an AFM tip on a substrate both made of diamond. The dimensions of the tip are estimates provided by QZabre company for the AFM tip we purchased. The excitation is a beam with wavelength $\lambda_0=800nm$ and Numerical aperture $NA=0.65$ which resembles the target beam in our experiment. Similar to the experimental setup, the beam is x-polarized before going through a QWP and travels through the fast axis of the wave plate when the wave plate angle ($\theta$) is zero. Due to reflection of the target beam from a dichroic mirror in the experiment, the polarization of the beam has an ellipticity of $\varepsilon=1.08$ before entering the objective lens when the QWP is at its optimal angle. However, this ellipticity is neglected in the simulations. The power of the beam is $P=4mW$ before the objective lens which has a measured transmission of $T=78\%$ at this wavelength. The power of the incident beam in the simulation is set to match this transmitted power in the experiment. The beam travels in $+\hat{z}$ direction and is focused at a distance $z_0=1.8\mu m$ if the medium was vacuum. The parameter $z_0$ is chosen such to qualitatively match the simulation results to the amplitude of the measured effective field.

Fig. \ref{fig:simulations}B shows the transverse effective field in the AFM tip in a plane $10nm$ away from the end of the tip. This plane coincides with the depth of the implanted NV center. The black circle shows the circumference of the diamond tip in this plane. The colorbar shows the amplitude of the transverse field and the black arrows show its direction. This transverse field gives rise to the DC offset in the experimental result (Fig. 3C). If the NV center is placed off-axes to the AFM tip there would be a non-zero transverse spin component at its location. This transverse component behaves differently from the longitudinal component of the photonic spin density when changing the QWP angle ({\it 41\/}), giving rise to a DC offset. 

To illustrate the effect of transverse spin, we show the effective magnetic field for an NV center in $\hat{n}=(\hat{x}+\hat{y}-\hat{z})/\sqrt{3}$ direction (same as the direction of the NV center in our experiment) placed at different locations on the transverse plane. Fig. \ref{fig:simulations}C shows the effective magnetic field along the direction of this NV center. Each curve corresponds to a different location for the NV center marked with white dots on Fig.~\ref{fig:simulations}B. The dots have a relative distance of $d=-10nm\hat{x}+10nm\hat{y}$ to each other. This figure clearly shows the development of the DC offset as the distance to the center of the tip increases. The Curve shown is Fig. 3C of the main text is the same as curve 7 on Fig. \ref{fig:simulations}C.

\pagebreak
{\bf References}
\begin{enumerate}
\addtocounter{enumi}{36}
\item W. Happer, B. S. Mathur, {\it Phys. Rev.} {\bf 163}, 12 (1967).
\item H. Deutsch, P. S. Jessen, {\it Phys. Rev. A} {\bf 57}, 1972 (1998).
\item J. R. Maze, et al., {\it New Journal of Physics} {\bf 13}, 025025 (2011).
\item Y. Chu, M. D. Lukin, {\it arXiv:1504.05990 [quant-ph]} (2015).
\item F. Kalhor, T. Thundat, Z. Jacob, {\it Applied Physics Letters\/}  {\bf 108}, 061102 (2016).
\end{enumerate}